\documentclass[11pt,a4wide]{article}
\usepackage{graphicx}
\usepackage{lscape}
\usepackage{color}
\usepackage[british]{babel}
\usepackage{amssymb,amsmath,amsfonts,verbatim,xkeyval,bm}
\usepackage{subfigure}
\usepackage{sidecap}
\usepackage{textcomp}
\usepackage[strict]{changepage}
\newcommand{\snot[1]}{{\scriptstyle \times 10}^{#1}}

\begin{document}
\begin{center}
{\Large\bf New Hamiltonian expansions adapted\\
to the Trojan problem}\\
\vskip 1cm
Roc\'{i}o Isabel P\'{a}ez$^1$,
Ugo Locatelli$^1$ and Christos\,Efthymiopoulos$^2$\\
$^1$Department of Mathematics, University of Rome "Tor Vergata"\\
$^2$Research Center for Astronomy and Applied Mathematics, Academy of Athens\\
\vskip 1cm
\end{center}

\noindent
{\small {\bf Abstract:} A number of studies, referring to the observed
  Trojan asteroids of various planets in our Solar System, or to
  hypothetical Trojan bodies in extrasolar planetary systems, have
  emphasized the importance of so-called {\it secondary resonances} in
  the problem of the long term stability of Trojan motions. Such
  resonances describe commensurabilities between the fast, synodic,
  and secular frequency of the Trojan body, and, possibly, additional
  slow frequencies produced by more than one perturbing bodies. The
  presence of secondary resonances sculpts the dynamical structure of
  the phase space. Hence, identifying their location is a relevant
  task for theoretical studies. In the present paper we combine the
  methods introduced in two recent
  papers~(\cite{PaezEfthy2015},~\cite{PaezLocat2015}) in order to
  analytically predict the location of secondary resonances in the
  Trojan problem. In~\cite{PaezEfthy2015}, the motion of a Trojan body
  was studied in the context of the planar Elliptic Restricted Three
  Body (ERTBP) or the planar Restricted Multi-Planet Problem
  (RMPP). It was shown that the Hamiltonian admits a generic
  decomposition $H=H_b+H_{sec}$. The term $H_b$, called the basic
  Hamiltonian, is a model of two degrees of freedom characterizing the
  short-period and synodic motions of a Trojan body.  Also, it yields
  a constant `proper eccentricity' allowing to define a third secular
  frequency connected to the body's perihelion precession.  $H_{sec}$
  contains all remaining secular perturbations due to the primary or
  to additional perturbing bodies. Here, we first investigate up to
  what extent the decomposition $H=H_b+H_{sec}$ provides a meaningful
  model.  To this end, we produce numerical examples of surfaces of
  section under $H_b$ and compare with those of the full model. We
  also discuss how secular perturbations alter the dynamics under
  $H_b$.  Secondly, we explore the normal form approach introduced
  in~\cite{PaezLocat2015} in order to find an `averaged over the fast
  angle' model derived from $H_b$, circumventing the problem of the
  series' limited convergence due to the collision singularity at the
  1:1 MMR. Finally, using this averaged model, we compute semi-analytically
  the position of the most important secondary resonances and compare
  the results with those found by numerical stability maps in specific
  examples.  We find a very good agreement between semi-analytical and
  numerical results in a domain whose border coincides with the
  transition to large-scale chaotic Trojan motions. }

\section{Introduction}\label{sec:intro}

Ever since the discovery of the triangular equilibrium solutions of
the Three Body Problem by Lagrange (1772), the problem of the
dynamical behavior of the orbits near the equilateral equilibrium
points has attracted great interest in the astronomical community. A
long known application refers to the family of Trojan asteroids of
Jupiter (see~\cite{RobSouc-10} and references therein). Trojan asteroids
were found also around other planets in our solar system, i.e. the
Earth, Mars, Uranus and
Neptune (\cite{Bowelletal-90},~\cite{Conetal-11},~\cite{Alexanetal-13}). On
different grounds, a number of works have adressed the questions of
the overall existence, formation and detectability of Trojan {\it
  exoplanets} (\cite{Beugetal-07},~\cite{CressNel-09}). No such body
has been identified so far in exoplanet surveys. This may indicate
that such planets are rare, which case would necessitate a dynamical
explanation, or that there exist yet unsurpassed constrains in
exo-Trojan detectability. It has
been proposed that the complexity of the orbits of Trojan bodies may
itself introduce intricacies in possible methods of detection, see,
for example,~\cite{Haghietal-13} refering to the Transit Timing
Variation method; regarding, in particular, the radial velocity
measurements, see~\cite{Dobro-13}.  The above and other examples
emphasize the need to understand in detail the orbital dynamics in the
1:1 Mean Motion commensurability.

In the present paper we extend the work of two previous 
papers (\cite{PaezLocat2015},~\cite{PaezEfthy2015}), 
in the direction of developing  
an efficient analytical method for the study of Trojan orbital dynamics. 
The aim of analytical studies is to identify the main features of the 
phase space and to quantify their role in the dynamical behavior of 
the orbits. Some important references of past analytical studies of 
the Trojan problem can be found in~\cite{Erdi-97} and references therein.

Regarding past approaches, the following is a key remark. Most
analytical treatments of the Trojan problem in the literature are so
far based upon series expansions of the equations of motion around the
stable equilibria $L_4$ and $L_5$, using various sets of variables
(e.g cartesian, cylindrical, or Delaunay-like action-angle
variables). However, it is important to recall that all these kinds of
expansions exhibit an important limitation, related to the singular
behavior of the equations of motion at relatively large Trojan
libration amplitudes. In the framework of the ERTBP, 
defined by a central mass, a perturber body and a massless particle
(the Trojan body), this singular behavior corresponds
geometrically to an approach of the Trojan body close to the
perturbing body, which is possible only at the 1:1 Mean Motion
Resonance. The relevant remark is that, the presence of a singularity
in the equations of motion implies a finite disc of convergence for
any kind of series expansions around $L_4$ or $L_5$. It is
straightforward to see that the projection of this disc in
configuration space is such so as to render the series' convergence
very poor for orbits with large libration amplitudes not only towards
the perturber, but also in the direction {\it opposite} to the 
perturber,
i.e. towards the unstable colinear point $L_3$. Let us note that this
poor convergence has a pure mathematical origin; no physical
singularity actually exists exactly at or close to $L_3$.

The following is a more precise form of the above remark. Let $\theta$ be 
the angular distance between the perturber and the Trojan body, e.g. in a 
heliocentric frame. Regardless the initial choice of variables, model 
approximation, etc., one finally recovers for $\theta$ a differential 
equation of the form (see, for example,~\cite{Erdi-78}) 
\begin{equation}\label{eqcri}
{d^2\theta\over dt^2} +  3\mu\sin\theta\left[1-2^{-3/2}
(1-\cos\theta)^{-3/2}\right]+\mbox{h.o.t.}=0~~
\end{equation}
where $\mu$ is the mass parameter of the perturber. The higher
order terms include epyciclic oscillations, the eccentricity of the
Trojan or the perturber, as well as any other kind of perturbation
induced, for example, by more perturbing bodies. Ignoring such terms,
Eq.~\eqref{eqcri} can be thought of as Newton's equation corresponding
to a `potential'
\begin{equation}\label{potpon}
V(\theta) = 3\mu\left[{1\over\sqrt{2-2\cos\theta}}-\cos\theta\right]~~. 
\end{equation}
This differs only by a constant from the quantity $H(\theta)$
introduced in~\cite{MurrDerm-99}, called also the `ponderomotive
potential' in~\cite{NamMurr-00}. If, instead, one expresses the
equations of motion in orbital elements, one encounters equivalent
terms in the {\it disturbing function} (\cite{MurrDerm-99} \S6),
taking the form $\mu[-\cos\tau+ (1-\cos\tau)^{-1/2}]$, where
$\tau=\lambda-\lambda'$ corresponds to the critical argument of the
1:1 Mean Motion commensurability, $\lambda$, $\lambda'$ being the mean
longitudes of the Trojan and the perturber respectively. The position
of $L_4$ (or $L_5$) corresponds to $\theta_0=\tau_0=\pi/3$ (or $5\pi/3$).
Setting $u=\theta-\theta_0$ or $u=\tau-\tau_0$ and expanding the
equations of motion in powers of the quantity $u$ leads to expressions
converging in the domain $|u|<\pi/3$. The convergence is quite slow
for angles approaching the limiting values $u_{lim}\pm\pi/3$. In
reality, such expansions become unpractical for libration angles $\sim
30^\circ$ and beyond, i.e. after half way to the singularity. The
applicability of all analytical methods based on polynomial expansions
around $L_4$ or $L_5$ is severely limited by this poor convergence.

\begin{figure}
  \centering 
  \includegraphics[width=.65\textwidth]{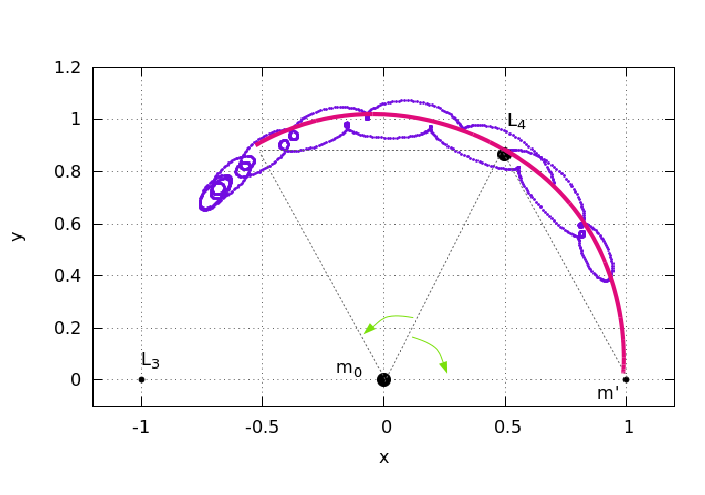}
  \caption[Representation of the convergence domain of series
    expansion around $L_4$]{ Representation of the domain of $\tau$
    where polynomial series are convergent if the expansion takes
    place around $L_4$, in a heliocentric cartesian frame
      co-rotating with the perturber $(x,y)$.  The position of the
    equilibrium points $L_4$ and $L_3$, the central mass $m_0$
    and the perturber $m'$ are indicated with black points.  The
    radius of convergence (thick pink line) of the series is given by
    the distance between $L_4$ and the perturber, namely
    $60^{\circ}$ ($\tau=\pi/3$).  While this does not induce any
    problem in the direction towards the perturber, it does
    limit the convergence in the opposite direction. In purple,
      we show an example of a typical Trojan orbit, obtained by
      numerically integrating the equations of motion of the ERTBP,
      for the initial condition $(x,y,\dot{x},\dot{y}) =
      (0.507,0.87402,0,0)$.  The orbit clearly exceeds the leftward
      limit of $60^{\circ}$ from $L_4$.}
\label{fig:convdomain1.png}
\end{figure}

On the other hand, one finds numerically that stable tadpole orbits 
exist in domains extending well beyond the limits of convergence of 
the analytical methods (see Fig. \ref{fig:convdomain1.png}). 

In~\cite{PaezLocat2015}, a new method of series expansions for the
Trojan problem was introduced, aiming, precisely, to remedy the poor
convergence of the classical series expansions around $L_4$ or $L_5$. The
method was developed in the context of the canonical formalism. In
more detail, an algorithm was derived allowing to compute a so-called
{\it Hamiltonian normal form} for Trojan motions. In the normal form
approach, starting from an initial Hamiltonian model, one performs a
series of near-identity canonical transformations from old to new
canonical variables, leading to a new expression for the Hamiltonian
(called the `normal form'). Via these transformations, the goal is to
arrive at a new form of the equations of motion in the new variables,
which is simpler to solve than in the original
variables. In~\cite{PaezLocat2015} the algorithm was applied to the
simplest possible model, namely the planar and circular Restricted
Three Body Problem (CRTBP). In this case, the normal form becomes an
{\it integrable} model of one degree of freedom, allowing to
analytically approximate the motion in the so-called synodic
(associated with the libration motion around $L_4$)  degree of
freedom. The key point of the method is that the functional dependence
of all involved quantities (i.e. normal form, transformation equations
etc., see section 3 below for details) on the quantity
\begin{equation}\label{b0}
 \beta_0={1\over\sqrt{1-\cos\tau}}
\end{equation}
and on the powers of $\beta_0$ is maintained at all orders of perturbation 
theory. Hence, the so-resulting series are not affected by the singularity 
at $\tau=0$ (i.e. $u=-\pi/3$) and remain useful practically within the whole 
tadpole domain. 

In the present paper we implement the method developed in~\cite{PaezLocat2015} 
in a model more realistic than the CRTBP, namely the model introduced 
in~\cite{PaezEfthy2015}. This provides an approximation to the Trojan 
dynamics applicable to two distinct cases: i) the planar Elliptic Restricted 
Three Body Problem (ERTBP), and ii) what was called in~\cite{PaezEfthy2015} 
the `Restricted Multi-Planet Problem' (RMPP). In the latter case, we assume 
that there are more than one perturbing bodies which exert secular 
perturbations on the Trojan body. The main application in mind is a 
hypothetical Trojan exoplanet in a multi-planet extrasolar system, although 
the model applies equally well to the Trojan asteroids of giant planets in 
our solar system. The RMPP exhibits a more rich spectrum of secular 
perturbations than the ERTBP. Even so, in~\cite{PaezEfthy2015} it was shown 
that in both problems, one can derive a so-called, `basic 
Hamiltonian model' (denoted hereafter as $H_b$). The Hamiltonian $H_b$ 
approximates the dynamics in the fast and synodic degrees of freedom. 
Furthermore, in~\cite{PaezEfthy2015} it was shown that $H_b$  
is formally identical in the ERTBP and the RMPP (apart from a re-interpretation 
of the physical meaning of one pair of action-angle variables). Consequently, 
the two problems are formally diversified only by their different sets of 
secular terms in the Hamiltonian, denoted by $H_{sec}$. Let us note that here, 
as in~\cite{PaezEfthy2015}, we focus only on the planar version of the $H_b$ 
model, although generalization to the spatial version is straightforward. 

Combining the results of~\cite{PaezEfthy2015}
and~\cite{PaezLocat2015}, we provide below an application of
particular interest, namely, the semi-analytical determination of the
location of {\it secondary resonances} in the tadpole domain of
motion. As was shown in~\cite{Erdietal-07}, secondary resonances play a key role
in determining the boundary and the size of the tadpole stability
domain.

In~\cite{PaezEfthy2015} a combination of numerical indicators (the
Fast Lyapunov Indicator - FLI~\cite{Froeschleetal-97}, as well as the
NAFF (Numerical Analysis of the Fundamental Frequencies)
algorithm~\cite{Laskar-04} were used to identify the most important
secondary resonances in a space corresponding to what is known as the
`proper elements' of the Trojan body's motion
(see~\cite{Milani-93},~\cite{BeauRoig-01}, as well as the definitions
in~\cite{PaezEfthy2015}). As an example, in the case of the ERTBP, the
location of various secondary resonances was determined in the space
of proper elements, depending mainly on two parameters, i.e., the
perturber's mass parameter $\mu$ and eccentricity $e'$.  In the present
paper, we demonstrate, instead, the efficiency of the analytical
normal form approach of~\cite{PaezLocat2015} in identifying the
location of secondary resonances in the space of proper elements. 

The structure of the paper is as follows: in Section 2 we examine some
features of the `basic Hamiltonian' model of~\cite{PaezEfthy2015}, and
validate the usefulness of the decomposition $H=H_b+H_{sec}$ by
performing a numerical exploration of the dynamics under $H_b$ alone,
as well as of how the latter compares to the full Hamiltonian
dynamics. Then, in Section 3 we implement the normal form method
introduced in~\cite{PaezLocat2015} to the Hamiltonian $H_b$, and check its
performance in the location of the secondary
resonances. Section 4 summarizes our conclusions.

\section{Basic Hamiltonian $H_b$: construction and features}\label{sec:Hb}

In this section, we aim to explore in some detail the features of the 
Hamiltonian model introduced in~\cite{PaezEfthy2015}, and to discuss the 
advantages and the limitations in the approximations of this model. 
For completeness, we begin by briefly reviewing the construction of 
the model. For more details, we defer the reader to~\cite{PaezEfthy2015}.

\subsection{Construction}\label{sec:constr_Hb}
In the framework of the planar elliptic Restricted Three Body Problem
(pERTBP), the equations of motion of the Trojan body ('massless body') 
depend on two physical parameters: i) the mass 
parameter $\mu=\frac{m'}{m'+M}$, where $M$ is the mass of 
the central mass and $m'$ the mass of the perturber
(also 'primary perturber' or simply 'primary'),
and ii) the eccentricity 
of the heliocentric orbit of the primary perturber, $e'$. In the ERTBP 
$e'$ and the major semi-axis of the primary's 
orbit are constant, set $a'=1$ in our units.

\begin{figure}
\centering
\includegraphics[width=.90\textwidth]{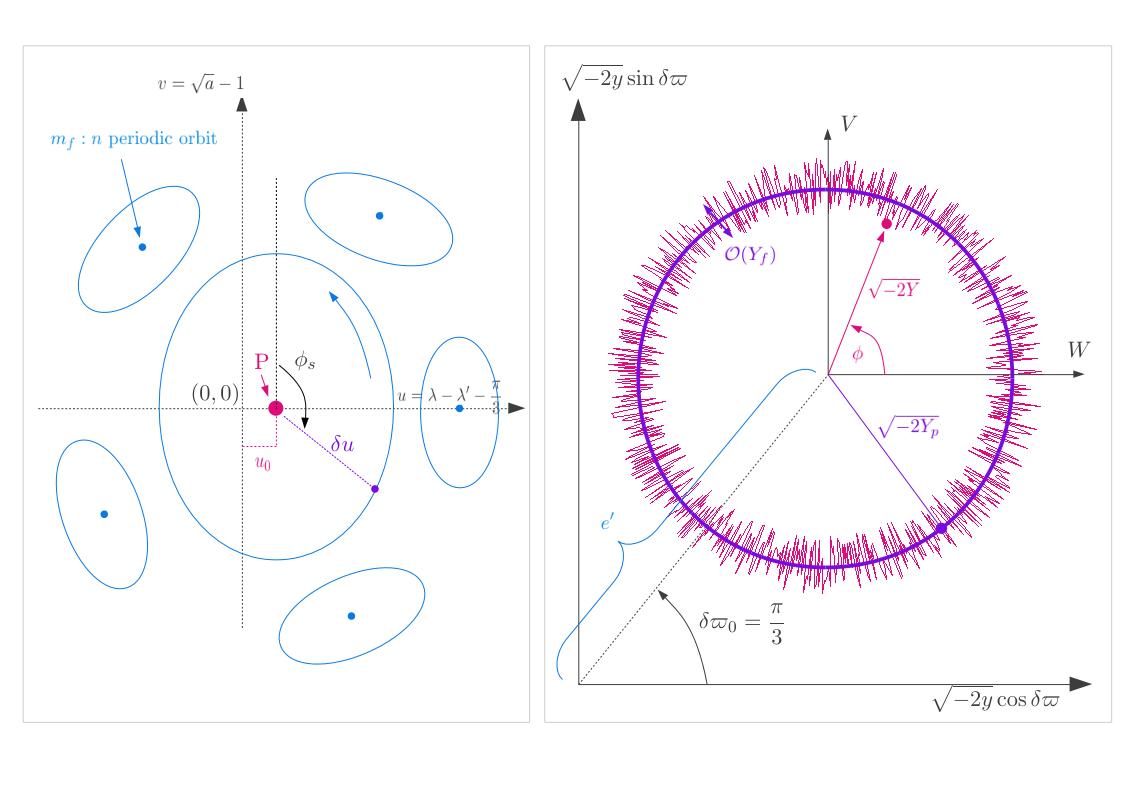}
\caption{{\footnotesize Schematic representation of the physical meaning of
    the action-angle variables used for the Hamiltonian $H_b$
    in~\eqref{eq:hbasic}. The plane $(u,v)$ corresponds to the
    `synodic' motion of the Trojan body, where $u=\lambda-\lambda'-
    \pi/3$ ($\lambda$ and $\lambda'$ correspond to the mean longitude
    of the Trojan body and the primary, respectively) and
    $v=\sqrt{a}-1$ ($a$ is the major semi-axis of the Trojan and $a'
    = 1$ for the primary).  Under the Hamiltonian $H_b$, the phase
    portrait can be represented by a Poincar\'{e} surface of section
    corresponding, e.g., to every time when the angle $\phi_f$
    accomplishes a full cycle. The left panel shows the form of the
    projection of this section on the plane $(u,v)$. The central point
    P represents a stable fixed point corresponding to the
    short-period periodic orbit around $L_4$. The orbit has frequency
    $\omega_f$, while its amplitude increases monotonically with
    $Y_f$. The forced equilibrium corresponds to $u_0=0$, $Y_f=0$. The
    point P, however, has in general a shift to positive values
    $u_0>0$ for proper eccentricities $e_p$ larger than zero. On the
    surface of section, the frequency of libration around the periodic
    orbit is given by the synodic frequency $\omega_s$. Resonances,
    and their island chains, correspond to rational relations between
    the fast frequency $\omega_f$ and $\omega_s$. On the other hand,
    the plane $(W,V)= (\sqrt{-2y}\cos\delta\varpi,\sqrt{-2y}\sin\delta\varpi,)$ 
    with $y = \sqrt{a}(\sqrt{1-e^2}-1)$ and $\delta \varpi = \varpi - \varpi'$
    the difference of longitude of perihelion of the Trojan and the
    primary (right panel) depicts the evolution of the Trojan body's
    eccentricity vector under the Hamiltonian $H_b$. The motion of the
    endpoint of the eccentricity vector can be decomposed to a
    circulation around the forced equilibrium, with angular frequency
    $g$, and a fast (of frequency $\omega_f$) `in-and-out' oscillation
    with respect to a circle of radius $e_p$, of amplitude which is of
    order ${\cal O}(Y_f)$. All extra terms with respect to $H_b$ in
    the Hamiltonian \eqref{eq:h_rmpp} depend on the slow angles
    $(\phi,\phi')$ in pERTBP, and also on the angles $\phi_j$,
    $j=1,\ldots,S$ in the RMPP. Thus, all these terms can only slowly
    modulate the dynamics under $H_b$.}}
  \label{fig:HBvar}
\end{figure}

In~\cite{PaezEfthy2015}, a Hamiltonian formulation was provided
  for the Trojan motion, in the pERTBP, and also in a more complex
  problem where $S$ additional perturbing bodies (e.g. planets) are
  present, being mutually far from MMRs. The Hamiltonian of the
  'Restricted Multi-Planet Problem' (RMPP) was written
  in~\cite{PaezEfthy2015} under the form
\begin{equation}\label{eq:h_rmpp}
\begin{aligned}
H &= H_b\,(Y_f,\phi_f,u,v,Y_p;\mu,e'_0) \\ 
&+\, H_{sec}\,(Y_f,\phi_f,u,v,Y_p,\phi,P',I_1,\ldots I_S,\phi',
\phi_1,\ldots,\phi_S)~~.
\end{aligned}
\end{equation}
In Eq.~\eqref{eq:h_rmpp}, the variables $(\phi_f,Y_f)$, $(u,v)$ and
$(\phi,Y_p)$ are pairs of action-angle variables, whose definition
stems from Delaunay-like variables following a sequence of four
consecutive canonical transformations (see Appendix).
The reader is deferred to Section 2 of~\cite{PaezEfthy2015} for the
details, while a schematic description of the physical meaning of these
variables is given in Fig.~\ref{fig:HBvar}. On the other hand, 
the angle $\phi'=g't$
corresponds to the longitude of the pericenter of the primary
perturber (constant in the pERTBP, precessing in the RMPP), 
while the angles $\phi_j=g_j t$ account for the secular
perturbations induced by the $S$ additional perturbers.  These angles
are canonically conjugate to a set of (dummy) action variables denoted
by $P'$ and $I_j$ respectively. Finally, $e_{0}'$ is the average 
eccentricity of the primary perturber, which coincides with $e'$ in
the pERTBP.

We call the term $H_b$ in the Hamiltonian of Eq.~\eqref{eq:h_rmpp} the `basic 
Hamiltonian model' for Trojan motions in the 1:1 MMR. Its detailed form 
is given in the Supplementary Online Material of~\cite{PaezEfthy2015}. We find
\begin{equation}\label{eq:hbasic}
H_b = -\frac{1}{2(1+v)^2}-v +(1+g')Y_f - g' Y_p - \mu {\cal F}^{(0)}
(u,\phi_f,v,Y_f-Y_p;e_0')~~.
\end{equation}
The function ${\cal F}^{(0)}$, contains terms depending on the canonical 
pairs $(\phi_f,Y_f)$ and $(u,v)$. The former characterizes fast motions 
(with frequency $\omega_f \sim {\cal O}(1)$), while the latter characterizes 
the `long-period' synodic motions (with frequency $\omega_s 
\sim {\cal O}(\sqrt{\mu})$). On the other hand, since the angle $\phi$ 
(see Fig.~\ref{fig:HBvar}) is ignorable in $H_b$, the action variable 
$Y_p$ is an integral of the basic Hamiltonian. This allows to define 
also a secular frequency via $g=\dot{\phi}=\partial H_b/\partial Y_p$. 
More precisely, we recover the well known relations (e.g., \cite{Erdi-88})
\begin{equation}\label{eq:fast_freq}
\omega_f \equiv \dot{\phi}_f = 1 - \frac{27}{8}\, \mu + g' + \ldots~~,
\end{equation}
\begin{equation}\label{eq:sin_freq}
\omega_s \equiv \dot{\phi}_s = - \sqrt{\frac{27 \mu}{4}} + \ldots ~~,
\end{equation}
\begin{equation}\label{eq:sec_freq}
g \equiv \dot{\phi} = \frac{27}{8}\, \mu - g' + \ldots~~.
\end{equation}
On the other hand, the higher order corrections in Eqs.~\eqref{eq:fast_freq},
\eqref{eq:sin_freq}, \eqref{eq:sec_freq}, can be recovered by an efficient
normal form approach, as shown in Section 3 below.

Three additional remarks concerning $H_b$ are: 

i) The constancy of $Y_p$ under $H_b$ allows to define an
approximation to the quasi-integral of the proper eccentricity $e_p$
(see~\cite{PaezEfthy2015}) via
\begin{equation}\label{eq:prop_ecc}
e_p = \sqrt{-2Y_p}~~.
\end{equation}
This approximation remains useful in the whole spectrum of models 
ranging from the CRTBP to the full RMPP. 

ii) In Eq.~\eqref{eq:hbasic}, the dependence of ${\cal F}^{(0)}$ on the 
actions $Y_p$ and $Y_f$ is exclusively via the difference $Y_f-Y_p$. 
This fact allows to simplify some normal form computations, as shown 
in Subsection \ref{sec:feat_Hb} below. We can define, in respect, 
an eccentricity parameter 
\begin{equation}\label{eq:fake_prop_ecc}
e_{p,0} = \sqrt{-2Y} = \sqrt{2Y_f-2Y_p}~~.
\end{equation}
The quantity $e_{p,0}$ will be used below in labeling several solutions 
found via the study of $H_b$.

iii) By construction, $H_{b}$ is formally identical in the RMPP and in 
the pERTBP, with the substitution $e'_0 \rightarrow e'$ and setting $g'=0$. 
Thus, the determination of the frequencies $\omega_f$, $\omega_s$ 
and $g$ based on a normal form manipulation of $H_b$ as below (Section 
\ref{sec:norm_Hb}) leads to equivalent results regardless the number of 
additional perturbing bodies besides the primary.  

On the other hand, $H_{sec}$ in \eqref{eq:h_rmpp} gathers all the terms 
of $H$ depending on the slow (secular) angle $\phi$ (with frequency 
$g \sim {\cal O}(\mu)$), or, in the case of the RMPP, also on the slow angles 
$\phi'$, $\phi_j$, $j=1,\ldots S$ (of frequencies ${\cal O}(\mu_j)$). 
As a consequence, in~\cite{PaezEfthy2015} it was proposed that the dynamics 
at secondary resonances can be approximated as a {\it slow modulation} of 
all the resonances produced by the basic model $H_b$, due to the additional 
influence of $H_{sec}$.  Considering the RMPP with $S$ bodies, the most 
general form of a planar secondary resonance is given by
\begin{equation}\label{eq:sec_reson_conmensu}
m_f \omega_f + m_s \omega_s + m g + m'g' + m_1 g_1 + \ldots + m_sg_s = 0~~,
\end{equation}
where $m_f$, $m_s$, $m$, $m'$, $m_j$ (with $j=1,\ldots,S$) are integers.
Keeping the notation of~\cite{PaezEfthy2015}, the most important secondary
resonances are those present in the basic Hamiltonian model $H_b$, already 
if $e'=0$, i.e. the resonances of the circular RTBP. These are denoted as 
the $m_f$:$m_s$ resonances, with comensurability relation
\begin{equation}\label{eq:main_sec_res}
m_f \omega_f + m_s \omega_s = 0 ~~.
\end{equation}
The particular case when $m_f =1$ corresponds to the lowest order
resonances that can be found for a certain value of the mass parameter
$\mu$, and usually dominate the structure of the phase-space. These 
are called the 'main secondary resonances' $1$:$n$, where $n=m_s$. 
For values of $\mu$ between $0.01$ and $0.0005$, $n$ corresponds to 
$4,5,6\ldots,16$.  On the other hand, we collectively refer to any 
other resonance of the ERTBP (involving all 3 frequencies $\omega_f$, 
$\omega_s$ and $g$) as well as to more general cases of the RMPP 
(including the frequencies $g'$, $g_j$) as 'transverse' resonances.

\subsection{Limits of applicability of the basic model 
$H_b$}\label{sec:feat_Hb}
The basic model $H_b$ represents a reduction of the number of degrees 
of freedom with respect to the original problem. Thus, we expect that 
its usefulness in approximating the full problem (ERTBP or RMPP) 
holds to some extent only.  The following numerical examples aim 
to compare the dynamical behavior of the orbits under the $H_b$ 
and the full Hamiltonian. To this end, we compute and compare 
various phase portraits (surfaces of section) arising under the two 
Hamiltonians. We restrict ourselves to the comparison between $H_b$ 
and the full Hamiltonian of the ERTBP only. We thus set $e_0'= e'$, 
and $g'=0$, $S=0$. Then, all secular perturbations are accounted for 
by only one additional degree of freedom with respect to $H_b$, 
represented by the canonical pair $(\phi,Y_p)$. Integrating 
numerically the RMPP instead of the ERTBP is considerably more 
expensive. Still, it is arguable that the effect of the secular 
perturbations should remain qualitatively similar by adding more  
degrees of freedom consisting of slow action-angle pairs only, 
as in the Hamiltonian decomposition of Eq.~\eqref{eq:h_rmpp}. 

Our numerical integrations of the full Hamiltonian model (ERTBP) are 
performed in heliocentric Cartesian variables, in which the equations 
of motion are straightforward to express. Whenever needed, translation 
from Cartesian to the canonical variables appearing in~\eqref{eq:h_rmpp} 
and vice versa is done following the sequence of canonical 
transformations defined in~\cite{PaezEfthy2015}. 

On the other hand, for the basic Hamiltonian $H_b$ we have an explicit
expression only in the latter variables. However, one can readily see
that, for fixed $(u,v,\phi_f)$, all the initial conditions of fixed
difference $Y_f-Y_p$ lead to the same orbit, independently of the
individual values of $Y_f$ or $Y_p$. If we set $Y_f=Y_{f,ref}=0$ and
$Y_p=Y_{p,ref}=-e_{p,ref}^2/2$ for one particular orbit chosen in
advance, that we call the `reference orbit', this allows to specify a
certain appropriate value of the energy $E=E_{ref}$ equal to the
numerical value of $H_b$ for that orbit. The reference orbit satisfies
the condition $e_{p,ref}=e_{p,0}$, i.e., $e_{p,0}$ becomes equal to
the modulus of the initial vector $\mathbf{e}-\mathbf{e}_{forced}$,
where $\mathbf{e}=(e\cos\omega,e\sin\omega)$, and $\mathbf{e}_{forced}
= (e'/2,e'\sqrt{3}/2)$. Now, keeping {\it both} $Y_p=Y_{p,ref}$ and
$E=E_{ref}$ fixed, but altering $(u,v,\phi_f)$, allows to solve the
equation $E_{ref}=H_b$ for $Y_f$ and specify new initial conditions
for more orbits at the same energy as the reference orbit. However,
now we will find in general that the initial value of $Y_f$ for any of
these new orbits satisfies $Y_f\neq 0$. In terms of the initial
eccentricity vector, this implies that $e_{p,0}\neq e_{p,ref}$.  The
so found orbit is the same as the one in which we set
$Y_p=-e_{p,0}^2/2\neq Y_{p,ref}$, and $Y_f=0$. For convenience, we
formally proceed with the former process (keeping $E=E_{ref}$ and
$Y_p=Y_{p,ref}$ fixed and adjusting $Y_f$ for different initial
conditions). However, since the value of the proper eccentricity for
each of these initial conditions $e_{p,0}$, we label all plots by
$e_{p,0}$ instead of $e_p$ in the FLI stability maps presented below
as well as in~\cite{PaezEfthy2015}.

Returning to our numerical computations, in order to choose a
reference orbit we select one close to the short period family around
$L_4$~\cite{Rabe-68}. More precisely, we set $u=v=\phi_f=Y_f=0$ for the
reference orbit, and consider different values for
$Y_p=Y_{p,ref}$. Physically, this means to choose different energy
levels $E=E_{ref}$ at which the reference orbit has different proper
eccentricity. Let us note that the existence of a central periodic
orbit is itself a property of the basic model $H_b$; adding more
degrees of freedom implies, instead, the existence of an invariant
torus of dimension larger than one and smaller than the full number of
degrees of freedom.

Having selected $E_{ref}$ and $Y_{p,ref}$, we compute initial
conditions for more orbits at the energy $E=E_{ref}$. More precisely,
in each of the figures which follow, we define a set of $19$ initial
conditions given by $u_j=0.05\times j$, $v_j=0$, $\phi_{f,j}=0$, for
$j=0,\ldots,18$, and $Y_{f,j}$ computed as described above. With these
initial conditions, we numerically integrate the orbits, under
the equations of motion of $H_b$, up to collecting, for each orbit,
500 points on the surface of section 
$\phi_f=0\,(\mathrm{mod}\:2\pi)$.

The same set of initial conditions is integrated under the equations 
of motion of the full ERTBP, for a time equivalent to 500 revolutions 
of the primary, collecting about 490 points in the same surface of 
section. In the ERTBP, the surface of section is four-dimensional, 
but a two-dimensional projection on the plane $(u,v)$ allows 
comparisons with the corresponding section of the basic model $H_b$. 

As an additional comparison, we also compute the surface of section 
provided by an intermediate model between the $H_b$ and the pERTBP.
We construct a 3 d.o.f Hamiltonian in the following way
\begin{equation}\label{eq:hbsec}
H_{b,sec} = H_b\,(Y_f,\phi_f,u,v,Y_p;\mu,e',e_{p,0}) + \langle F^{(1)} \rangle
(u,v,Y_p,\phi;\mu,e',e_{p,0},Y_f)~~,
\end{equation}
where 
\begin{equation}\label{eq:f1ave}
\langle F^{(1)} \rangle={1\over 2\pi}\int_0^{2\pi}H_{sec}d\phi_f~~.
\end{equation}
Explicit formulae for $\langle F^{(1)} \rangle$ can be found in the
Supplementary Online Material of~\cite{PaezEfthy2015}. Such terms
may depend on the slow angle $\phi$, but are independent 
of the fast angle $\phi_f$.  Hence, $H_{b,sec}$ contains some, but not all,
the secular terms of the disturbing function of the pERTBP. 
On the other hand, up to first order 
in the mass parameter $\mu$, the averaging~\eqref{eq:f1ave} yields the 
same Hamiltonian as the one produced by a canonical transformation 
eliminating all terms depending on the fast angle $\phi_f$. 
Thus, the model $H_{b,sec}$ captures the main effect of the secular 
terms, as discussed in~\cite{PaezEfthy2015}, which is a \emph{pulsation},
with frequency $g$, of the separatrices of all the secondary resonances 
induced by $H_b$. Since the modulation due to these secular terms is 
slow, far from secondary resonances we expect that an adiabatic invariant 
holds for initial conditions close to the invariant tori of $H_b$, 
thus yielding stable regular orbits. On the other hand, in~\cite{PaezEfthy2015}
it was argued that close to secondary resonances the pulsation provokes 
a weak chaotic diffusion best described by the paradigm of modulational 
diffusion. 

\begin{figure}
  \centering
  \includegraphics[width=0.75\textwidth]{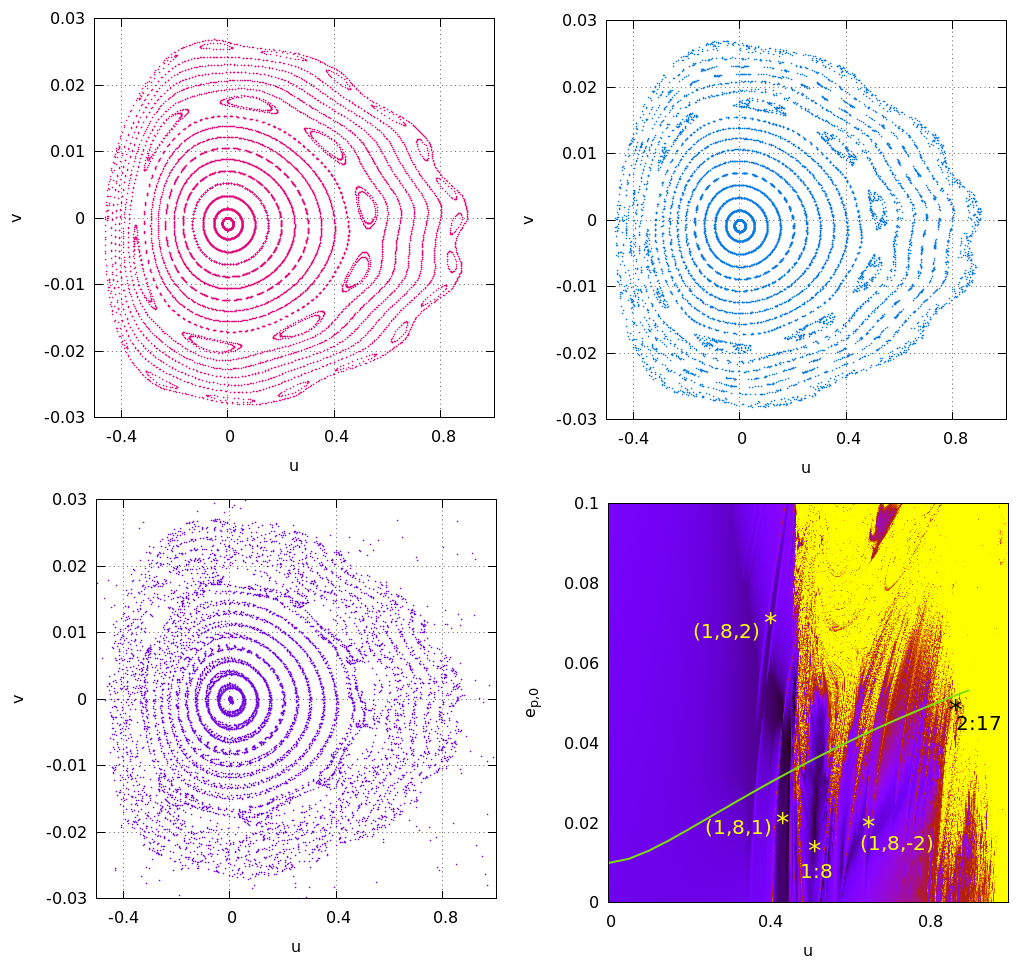}
  \caption{Comparison of surfaces of section (section condition
    $\phi_f=0$) provided by different models.  The considered
    parameters are $\mu=0.0024$, $e'=0.04$ and $e_{p,ref}=0.01$. In
    pink points (upper left), we show the surface of section provided
    by $H_b$. In blue points (upper right), the one corresponding to
    $H_{b,sec}$. In purple points (lower left), the one corresponding
    to pERTBP. In lower right panel, we reproduce the FLI map
    of~\cite{PaezEfthy2015} corresponding to the physical parameters
    $\mu$ and $e'$ considered, with the most important secondary
    resonances indicated. The color-scale for the FLI map goes as
      follows: dark colors (purple) indicate regular orbits, while
      light colors (yellow) indicate for the chaotic orbits
      (see~\cite{PaezEfthy2015} for the exact FLI computation). The
    green line on the FLI map indicates the isoenergetic curve where
    the initial conditions are located.}
  \label{fig:plot-epr01}
\end{figure}

\begin{figure}
  \centering
  \includegraphics[width=0.75\textwidth]{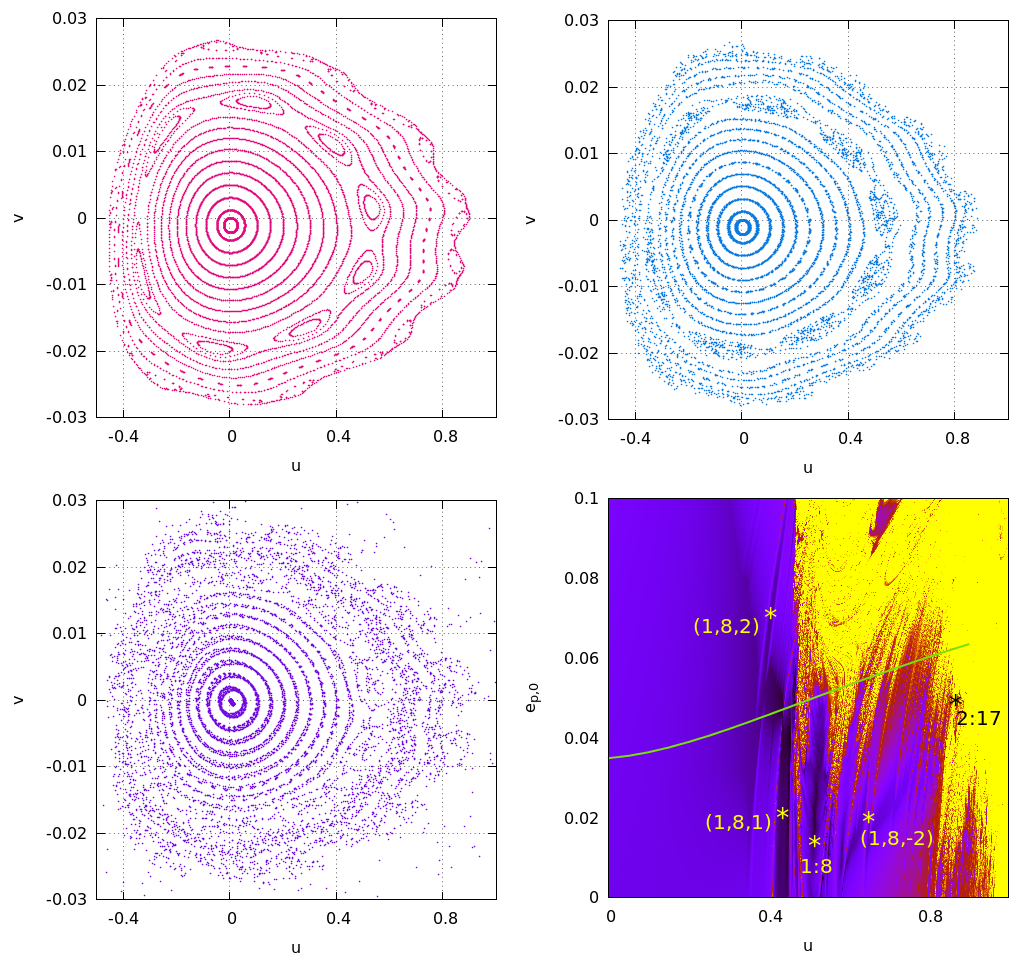}
  \caption{Same as in Fig.~\ref{fig:plot-epr01}, but for a higher parameter 
    value $e_{p,ref}=0.035$. }
  \label{fig:plot-epr035}
\end{figure}

\begin{figure}
  \centering
  \includegraphics[width=0.75\textwidth]{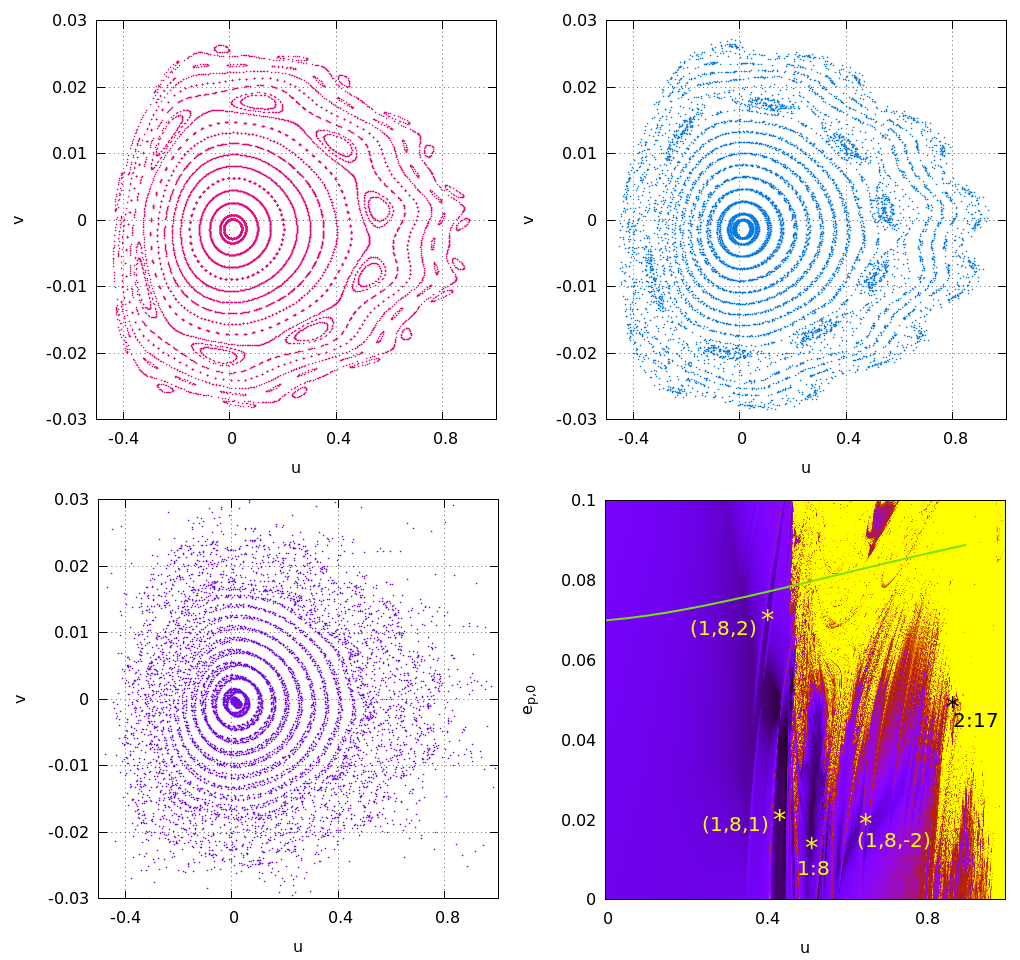}
  \caption{Same as in Fig.~\ref{fig:plot-epr01}, but for a still higher 
    parameter value $e_{p,ref}=0.07$.}
  \label{fig:plot-epr07}
\end{figure}

Figures~\ref{fig:plot-epr01},~\ref{fig:plot-epr035}
and~\ref{fig:plot-epr07} show an example of the comparison between the three 
mentioned above models. The physical parameters chosen for these 
plots are $\mu=0.0024$ (which depicts clearly the 1:8 main secondary resonance) 
and $e'=0.04$. Figure~\ref{fig:plot-epr01} shows the surface of section
$\phi_f = 0\,(\mathrm{mod}\:2\pi)$ 
corresponding to $e_{p,ref}=0.01$, Fig.~\ref{fig:plot-epr035} to 
$e_{p,ref}=0.035$ and Fig.~\ref{fig:plot-epr07} to $e_{p,ref}=0.07$. 
In each figure, the upper left plot (pink points) corresponds to the surface 
of section produced by the flow under the basic model $H_b$, the upper right 
plot (blue points) to the flow under $H_{b,sec}$ and the lower left plot 
(purple points) to the flow under the full Hamiltonian of the pERTBP. 
As an additional information, we provide the FLI stability map corresponding 
to the same parameters $\mu$ and $e'$, which was computed in Fig. 8c
of~\cite{PaezEfthy2015} (see that paper for details on the FLI computation). 
On top of the FLI map, in green we show the locus of initial conditions 
$(u,e_{p,0})$ on the surface of section whose orbits have constant energy 
$E=E_{ref}$. 

In Fig.~\ref{fig:plot-epr01}, in the approximation based on the model
$H_b$, the absence of any dependence of the dynamics on the slow angle
$\phi$ renders possible to clearly display the short period and
synodic dynamics by means of the surface of section
$\phi_f = 0\,(\mathrm{mod}\:2\pi)$, which, for $H_b$,
is two-dimensional. In fact, for more complex models like $H_{b,sec}$
or the full pERTBP, the corresponding surface of section is
4-dimensional and its 2D projection on the $(u,v)$ plane becomes
blurred (top right and bottom left panels respectively).  The blurring
can be due partly to projection effects. However, we argue below that
an important effect is caused also by the influence of the secular
terms, absent in $H_b$, to the dynamics.

Returning to the phase portrait of $H_b$, this allows to extract
relevant information such as: i) the position of the central fixed
point, corresponding to the crossing of the section by the short
period orbit, ii) several secondary resonances and the corresponding
resonant islands of stability, and iii) the overall size of the
libration domain of \emph{effective} stability. Also, this phase
portrait allows to understand the structure of the stability map. In
the phase portrait, as we move from left to right along the line
$x=0$, we encounter non-resonant tori, interrupted by thin chaotic
layers and the islands of some secondary resonances, namely the
resonances 1:8 (at $u \sim 0.55$) and 2:17 (at $u \sim 0.85$). 
Note, however, that no transverse secondary resonances can be
seen in the $H_b$ portrait, since these resonances correspond, in
general, to a non-resonant frequency ratio of the fast and synodic
frequencies $\omega_f$ and $\omega_s$; except at resonance junctions,
the exact resonance condition $m_f\omega_f + m_s\omega_s +
m_g g=0$ for some non-zero $m_s$, $m_f$, $m_g$ implies, in
general, non-commensurable values of $\omega_f$ and $\omega_s$. Since
$q \ll \omega_s \ll \omega_f$, most transverse resonances can only
accumulate close to the main secondary resonances forming resonant
multiplets, as confirmed by visual inspection of the stability maps
(see also~\cite{PaezEfthy2015}). However, some isolated transverse
resonances may be embedded in the main domain of stability whose
border is marked by the most conspicuous secondary resonance.  In
Fig.~\ref{fig:plot-epr01}, this domain extends up to about $u\approx
0.5$. In the stability map of Fig.~\ref{fig:plot-epr01}, the
transverse resonances $[1,-8,k]$, with $k=-2,-1,1,2$ form a multiplet
together with the conspicuous resonance $1$:$8$. Two of these
transverse resonances ($k=2$ and $k=1$) are embedded in the main
domain of stability. However, none of the transverse resonances is
visible in the phase portrait of the basic model $H_b$.

We now discuss the pulsation effect of the phase portrait due to the
slow modulation induced by the secular terms. As shown
in~\cite{PaezEfthy2015}, the amplitude of the secular terms depends on
the values of $e'$ and $e_{p,0}$.  For fixed $e'\neq 0$, the amplitude
of the pulsation generated by such terms increases with $e_{p,0}$. For
values of $e_{p,0}$ large enough, the pulsation modifies the whole
behavior in phase-space. Since, along the line $x=0$, $e_{p,0}$
increases with $u$ (green curve in low-right panel of
Fig.~\ref{fig:plot-epr01}), the amplitude of the pulsation increases
as we move from the central fixed point outwards. In regions where the
resonant web is dense enough, this pulsation causes all narrow
transverse resonances in a multiplet to overlap, increasing the size
of the chaotic domain and facilitating escaping mechanisms.  For the
set of parameters of Fig.~\ref{fig:plot-epr01}, we see from the
corresponding FLI map that this happens for values of $e_{p,0}$
greater than about $0.06$.  Beyond this value, the effect induced by
$H_{sec}$ implies that the blurring observed in the phase portraits
(apart from the one of $H_b$) is not due just to projection effects
but it has a dynamical origin, the nature of the orbits changes as
they are converted from regular to chaotic. Evidence of this
  phenomenon is found, e.g, in the case of the resonance 2:17. While in
  the surface of section of the $H_b$, the 2:17 stability islands are
  clearly seen, such resonance is not evident in the surfaces of
  section of the ERTBP and $H_{b,sec}$. As represented by the FLI map,
  the effect of the resonance's separatrix pulsation results in that
  no libration domain is identifiable in the FLI map.

This latter effect is more conspicuous in Figs. \ref{fig:plot-epr035} and 
\ref{fig:plot-epr07}, in which, choosing a higher $e_{p,ref}$, we increase 
the level of proper eccentricities of all the orbits. 
In Fig.~\ref{fig:plot-epr035}, the FLI stability map shows large domains 
of chaos which are not observed in the phase portrait of $H_b$, but they 
appear in the phase portrait of the full model. The separatrix pulsation 
of the 1:8 resonance is not, however, large enough so as to completely 
wash out this resonance, which is therefore seen in all four panels of 
the plot. On the other hand, increasing still more the level of proper 
eccentricities (Fig.~\ref{fig:plot-epr07}) makes this pulsation large 
enough so as to completely introduce chaos in the position of the 1:8 
resonance. This limit of eccentricity levels marks the overall validity 
of the approximation based on $H_b$ regarding the position of secondary 
resonances. Beyond this value, $H_b$ still represents fairly well the 
dynamics only inside the main librational domain of stability. We note 
also that the elimination of the main secondary resonance 1:8 by the 
separatrix pulsation is already present in the model $H_{b,sec}$ 
(compare the corresponding phase portraits in the three Figures 
\ref{fig:plot-epr01}, \ref{fig:plot-epr035}, \ref{fig:plot-epr07}). 

In conclusion, the pulsation mechanism induced by the secular terms in
the Hamiltonian affects essentially the regions of the phase space
where resonances accumulate in the form of multiplets. For libration
orbits, these are the regions beyond the main secondary resonance
$1$:$n$, which always dominates the phase-space.  The regions inner to
that resonance are not influenced considerably and the representation
of the dynamics via the basic model $H_b$ remains accurate there, even
for high values of the proper eccentricity. The value of the latter at
which the separatrix pulsation of the $1$:$n$ resonance completely
washes this resonance marks the overall limit of approximation of the
basic model. On the other hand, most orbits beyond that limit turn to
be chaotic and fast-escaping the libration domain, thus of lesser
interest in applications related to Trojan or exo-Trojan objects.

\section{Normal form}\label{sec:norm_Hb}

In~\cite{PaezLocat2015}, a new normalizing scheme was introduced for
the Hamiltonian of the planar Circular Restricted Three Body Problem
(pCRTBP).  Here, we adapt the scheme in order to compute a normal form
in the case of the basic model $H_b$ derived from the pERTBP. The
particular application considered is the semi-analytic determination of the
position of the secondary resonances in the plane of the Trojan body's
proper elements.

\subsection{Hamiltonian preparation}\label{sec:preparation}
The novelty of the normalizing scheme introduced
in~\cite{PaezLocat2015} lies on the way the scheme deals with the
synodic degree of freedom, expressed in the Hamiltonian through the
variables $(u,v)$. For obtaining the dynamics in the synodic variables
via a normal form, it is only necessary to average the Hamiltonian
$H_b$ over the fast angle $\phi_f$. The novelty consists of retaining
the original non-polynomial and non-trigonometric-polynomial
functional dependence of the Hamiltonian on the synodic angle $u$ in
all normal form expansions.  As pointed out in the introduction, this
allows to deal efficiently with the model's singular behavior at
$u=-\pi/3$.

We start by first expressing the basic model $H_b$ in variables 
appropriate for introducing the normalization scheme of~\cite{PaezLocat2015}. 
The synodic degree of freedom is represented 
by the variables
\begin{equation}\label{eq:shift_cent}
v=x-x_0, \quad u=\tau-\tau_0~~,
\end{equation}
where 
\begin{equation}\label{eq:syn_dof}
x = \sqrt{a}-1 \,,\quad\, \tau = \lambda-\lambda'~~,
\end{equation}
$a$ being the major semi-axis of the Trojan body, and $\lambda$,
$\lambda'$ the mean longitudes of the Trojan body and the primary
respectively.  The constants $x_0$ and $\tau_0$ in
\eqref{eq:shift_cent} give the position of the forced equilibrium of
the Hamiltonian averaged over $\lambda'$
(see~\cite{PaezEfthy2015}). In the case of the pERTBP, in the vicinity
of $L_4$, we have $x_0=0$, $\tau_0=\pi/3$. Finally, it turns
convenient to introduce new canonical pairs: $(\phi_f,~{\cal
  Y}=Y_f-Y_p)$, and $(\theta=\phi+\phi_f,~Y_p)$.  After these
preliminary transformations, the basic model $H_b$ reads
\begin{equation}\label{eq:hbasic_xtau}
H_b = -\frac{1}{2(1+x)^2}-x + {\cal Y}+Y_p
- \mu {\cal F}^{(0)}(\tau,\phi_f,x,{\cal Y};e')~~.
\end{equation}
The dependence of $H_b$ on $\tau$ is of the form $\frac{\cos^{k_1} \tau}
{(2-2\cos\tau)^{j/2}}$ or $\frac{\cos^{k_2} \tau}{(2-2\cos\tau)^{j/2}}$, 
$j=2n-1$ with $k_1$, $k_2$ and $n$ integers (see Supplementary Online Material
of~\cite{PaezEfthy2015}). Also, since the angle $\theta$ is ignorable,
$Y_p$ is a constant that can be viewed as a parameter in $H_b$.

In order to initialize the normalization procedure, we write and
expand the Hamiltonian in \eqref{eq:hbasic_xtau}, by introducing
modified Delaunay-Poincar\'{e} variables, as in~\cite{PaezLocat2015}
\begin{equation}\label{eq:delau-garf-coord}
\begin{aligned}
x &\,,\qquad \tau\,\\
\xi &=\sqrt{2{\cal Y}}\cos\phi_f\,,\\
\eta&=\sqrt{2{\cal Y}}\sin\phi_f\,~~.
\end{aligned}
\end{equation}
The new expression for the Hamiltonian reads
\begin{equation}\label{eq:hbasic_xtauxieta}
H_b(\tau,x,\xi,\eta,Y_p) = -\frac{1}{2(1+x)^2}-x + Y_p +
\frac{\xi^2+\eta^2}{2} - \mu {\cal F}^{(0)}
(\tau,x,\xi,\eta;Y_p,e')~~.
\end{equation}
Finally, we expand the Hamiltonian in terms of every variable except
$\tau$, obtaining

\begin{align}
H_b (\tau,x,\xi,\eta,Y_p) =& -x + \sum_{i=0}^{\infty}
\,(-1)^{i-1}(i+1)\, \frac{x^i}{2}\, +\, \frac{\xi^2+\eta^2}{2} \,+
\,Y_p \label{eq:hb_xtauxieta_exp}\\ +& \,\mu \sum_{\substack{
    m_1,m_2,m_3\\ k_1,k_2,k_3,j}} a_{m_1,m_2,m_3,k_1,k_2,j}\, e'^{k_3}
x^{m_1} \, \xi^{m_2}\, \eta^{m_3} \, \cos^{k_1}(\tau) \,
\sin^{k_2}(\tau) \, \beta^j(\tau)~~, \nonumber
\end{align}
where $a_{m_1,m_2,m_3,k_1,k_2,j}$ is a rational number and 
$\beta(\tau)=\frac{1}{\sqrt{2-2\cos\tau}}$. The Hamiltonian $H_b$ 
in~\eqref{eq:hb_xtauxieta_exp} represents the `normal form at the 
zero-th step in the normalizing scheme', i.e., before any normalization. 
This we denote as $H^{(1,0)}$.

\section{Normalizing scheme}\label{sec:norm_sche}

The normalizing algorithm defines a sequence of Hamiltonians by an
iterative procedure. In order to simplify some of the concepts below
we define the class ${\cal P}_{s,l}$ as the set of functions whose expansion
is of the form
\begin{equation}\label{eq:classP}
\sum_{2m_1+m_2+m_3=l}\, \, \sum_{\substack{k_1+k_2\leq l+4s-3\\ j\leq 2l+7s-6}} 
a_{m_1,m_2,m_3,k_1,k_2,j}\, e'^{k_3} x^{m_1} \, \xi^{m_2}\, \eta^{m_3}
\, \cos^{k_1}(\tau) \, \sin^{k_2}(\tau) \,\beta^j(\tau)~~.
\end{equation}
Let $r_1,\,r_2$ be two integer counters, $1 \leq r_1 \leq R_1$ and $1
\leq r_2 \leq R_2$ with fixed $R_1,R_2\in \mathbb{N}$. We assume that at 
a generic normalizing step ($r_1$,$r_2-1$), the expansion of the Hamiltonian 
is given by
\begin{equation}\label{eq:hr1r2-1}
\begin{aligned}
H^{(r_1,r_2-1)}(x,\xi,\tau,\eta,Y_p) = &\, Y_p+\frac{\xi^2 + \eta^2}{2} +
\sum_{i=2}^{\infty} \alpha_i\, x^i  \\
+ &\, \sum_{s=1}^{r_1-1}  \sum_{l=0}^{R_2} \mu^s
 Z_{s,l} \,( x, (\xi^2+\eta^2)/2,\tau )  \\
+ &\, \sum_{l=0}^{r_2-1} \mu^{r_1} Z_{r_1,l}\, (x, (\xi^2+\eta^2)/2,\tau)\\
+ &\, {\cal R}^{(r_1,r_2-1)} (x,\xi,\eta,\tau)~,
\end{aligned}
\end{equation}
where $\alpha_i$ are real coefficients and the remainder
${\cal R}^{(r_1,r_2-1)}(x,\xi,\eta,\tau)$ is given by
\begin{equation}\label{eq:Rr1-1r2-1}
\begin{aligned}
  {\cal R}^{(r_1,r_2-1)} &\,(x,\xi,\eta,\tau) =
   \mu^{r_1} f_{r_1,r_2}^{(r_1,r_2-1)}(x,\xi,\eta,\tau) + 
   \sum_{l=r_2+1}^{R_2} \mu^{r_1} f_{r_1,l}^{(r_1,r_2-1)}(x,\xi,\eta,\tau) \\
+ &\, \sum_{s=r_1+1}^{\infty} 
\sum_{l=0}^{R_2} \mu^s f_{s,l}^{(r_1,r_2-1)} (x,\xi,\eta,\tau) +
 \sum_{s=1}^{\infty} \sum_{r=R_2+1}^{\infty} \mu^s 
f_{s,l}^{(r_1,r_2-1)}(x,\xi,\eta,\tau)~.
\end{aligned}
\end{equation}

All the terms $Z_{s,l}$ and $f_{s,l}^{(r_1,r_2-1)}$ appearing
in~\eqref{eq:hr1r2-1} are made by expansions including a \emph{finite}
number of monomials of the type given by the class ${\cal
  P}_{s,l}$. More specifically $Z_{s,l} \in {\cal P}_{s,l}$
$\forall\ 0\le l\le R_2\,,\ 1\le s<r_1\,$, $Z_{r_1,l}\in {\cal
  P}_{r_1,l}$ $\forall\ 0\le l<r_2\,$, $f_{r_1,l}^{(r_1,r_2-1)}\in
{\cal P}_{r_1,l}$ $\forall\ l\ge r_2\,$, $f_{s,l}^{(r_1,r_2-1)}\in
{\cal P}_{s,l}$ $\forall\ l>R_2\,,\ 1\leq s<r_1\,$ and $\forall\ l\ge
0,\ s>r_1\,$.

In formula~\eqref{eq:hr1r2-1}, one can distinguish the terms in normal form
from the remainder ${\cal R}$: the latter depend on $(\xi,\eta)$ in
a generic way, while in the normal form terms $Z$, those variables just appear
under the form $(\xi^2+\eta^2)/2$.  The $(r_1,r_2)$--th step of the
algorithm formally defines the new Hamiltonian $H^{(r_1,r_2)}$ by
applying the Lie series operator $\exp {\cal L}_{\mu^{r_1}
\chi_{r_1,r_2}}$ to the previous Hamiltonian $H^{(r_1,r_2-1)}$, 
as it follows%
\footnote{We stress here that after each transformation we do not 
change the name of the canonical variables in order to simplify the 
notation.}
\begin{equation}\label{eq:hr1r2Lie}
H^{(r_1,r_2)} = \exp \left({\cal L}_{\mu^{r_1}\chi_{r_1,r_2}}\right) 
H^{(r_1,r_2-1)}~~.
\end{equation}
The Lie series operator is given by
\begin{equation}\label{eq:Lie_oper}
\exp \left({\cal L}_{\chi} \right) \, \cdot \, = \sum_{j \geq 0} \frac{1}{j!} 
{\cal L}_{\chi}^{j} \; \cdot~~,
\end{equation}
where the Lie derivative ${\cal L}_{\chi} g = \{ g, \chi \}$, is such that
$\{\cdot,\cdot\}$ is the classical Poisson bracket.
The new generating function $\mu^{r_1}\chi_{r_1,r_2}$ is
determined by solving the following homological equation with respect
to the unknown
$\chi_{r_1,r_2}= \chi_{r_1,r_2}(x,\xi,\tau,\eta)$:
\begin{equation}\label{eq:homol_eq}
{\cal L}_{\mu^{r_1} \chi_{r_1,r_2}} Z_{0,2} +
f_{r_1,r_2}^{(r_1,r_2-1)} = Z_{r_1,r_2} ~~,
\end{equation}
where $Z_{0,2} = \frac{\xi^2 +\eta^2}{2}$ and $Z_{r_1,r_2}$ is the
new term in the normal form, i.e. $Z_{r_1,r_2} = Z_{r_1,r_2}
(x,\tau,(\xi^2+\eta^2)/2)$. In other words,
$\mu^{r_1}\chi_{r_1,r_2}$ is determined so as to remove the terms
that do \emph{not} belong to the normal form from the main perturbing
term $\mu^{r_1} f_{r_1,r_2}^{(r_1,r_2-1)}$. Thus, by construction,
the new Hamiltonian $H^{(r_1,r_2)}$ inherits the structure of
Eq.~\eqref{eq:hr1r2-1}.  From the latter, we point out that the
splitting of the Hamiltonian in sub-functions of the form ${\cal
  P}_{s,l}$, organizes the terms in groups with the same order of
magnitude $\mu^s$ and total degree $l/2$ (possibly semi-odd) in the
variables $x$ and ${\cal Y}=\frac{\xi^2+\eta^2}{2}$. This way, we
exploit the existence of the natural small parameters of the model in
the normalizing procedure.  Furthermore, after having omitted the
constant term $\alpha_0\,$, we
can set the Hamiltonian $H_b$ in~\eqref{eq:hb_xtauxieta_exp} as the
first normalizing step Hamiltonian $H^{(1,0)}$, according
to~\eqref{eq:hr1r2-1}.

The algorithm requires just $R_1\cdot R_2$ normalization steps, 
constructing the finite sequence of Hamiltonians
\begin{equation}\label{eq:theHs}
H^{(1,0)} = H_b, \, H^{(1,1)},\, \ldots,\,H^{(1,R_2)},\, 
H^{(2,1)},\, \ldots,\, H^{(R_1,R_2)}~~.
\end{equation}
Here, we add the prescription that $H^{(r_1,0)} = H^{(r_1-1,R_2)}\, \forall \, 
1 < r_1 \leq R_1$.
Then, we write the final Hamiltonian, where we distinguish the normal 
form part from the remainder, as
\begin{equation}\label{eq:HR1R2}
H^{(R_1,R_2)}(x,\xi,\tau,\eta,Y_p) = {\cal Z}^{(R_1,R_2)} 
\left(x,\frac{(\xi^2+\eta^2)}{2}, \tau,Y_p \right) 
+ {\cal R}^{(R_1,R_2)} (x,\xi,\tau,\eta)~~. 
\end{equation}

At this point, we must remark a few features of the normal form ${\cal
  Z}^{(R_1,R_2)}$. While its dependence on $x$ and $\tau$ remains
generic, it depends on $\xi$ and $\eta$ \emph{only} through the form
$\frac{\xi^2+\eta^2}{2}$.  That is, we have
\begin{equation}\label{eq:HR1R2renamed}
H^{(R_1,R_2)}(x,\tau,{\cal Y},\phi_f,Y_p) = 
{\cal Z}^{(R_1,R_2)}\left(x,\tau,{\cal Y},Y_p\right) 
+ {\cal R}^{(R_1,R_2)} (x,\tau,{\cal Y},\phi_f)~~. 
\end{equation}
The key remark is that $\phi_f$ becomes ignorable in the normal form, 
and, therefore, ${\cal Y}$ becomes an integral of motion of 
${\cal Z}^{(R_1,R_2)}$. Then, the normal form can be viewed as a 
Hamiltonian of one degree of freedom depending on two constant actions 
${\cal Y}$ and $Y_p$, i.e. ${\cal Z}^{(R_1,R_2)}$ represents now a formally 
\emph{integrable} dynamical system.  Of course, since the true system 
is not integrable, it is natural to expect that the normalization 
procedure diverges in the limit of $R_1,R_2\rightarrow \infty$. 
The divergence corresponds formally to the fact that the size of the 
remainder function ${\cal R}^{(R_1,R_2)}$ cannot be reduced to zero 
as the normalization order tends to infinity. Then, the {\it optimal} 
normal form approximation corresponds to choosing the values of both 
integer parameters $R_1$ and $R_2$ so as to reduce the size of the 
remainder ${\cal R}^{(R_1,R_2)}$ as much as possible. In practice, 
there are computational limits that compromise the choice of values 
of $R_1$ and $R_2$. In all subsequent computations, the values are 
$R_1=2$ and $R_2=4$, corresponding to a second order expansion and 
truncation on the mass parameter $\mu$ and fourth order for the total polinomial
degree of $x$, $\xi$ and $\eta$. These normalization orders prove to be 
sufficient for the normal form to represent a good representation 
of the original Hamiltonian in the domain of regular motions. 
In particular, we will now employ this possibility in order to 
compute the positions of different secondary resonances, based on the 
integrable approximation provided by our normal form. 

\subsection{Application: determination of the location of resonances 
via the normal form} 
\label{sec:location}
Consider an orbit with initial conditions specified in terms of the two 
parameters $u=\tau-\tau_0$ and $e_{p,0}$ in the same way as in the 
stability maps of Figures \ref{fig:plot-epr01}, 
\ref{fig:plot-epr035}, \ref{fig:plot-epr07}. We will make use of the 
normal form approximation $Z^{(R_1,R_2)}$ in~\eqref{eq:HR1R2renamed} 
in order to compute the values of the three main frequencies of motion 
for the given initial conditions. The computation proceeds by the 
following steps:

1) We first evaluate the synodic frequency $\omega_s$, i.e., the 
frequency of libration of the synodic variables $\tau$ and $x$. 
The normal form $Z^{(R_1,R_2)}$ leads to Hamilton's equations:
\begin{equation}\label{eq:xdot}
\frac{\textrm{d}x}{\textrm{d}t} = f (x,\tau;{\cal Y}) =
-\frac{\partial Z^{(R_1,R_2)}}{\partial \tau}~~
\end{equation}
and 
\begin{equation}\label{eq:taudot}
\frac{\textrm{d}\tau}{\textrm{d}t} = g (x,\tau;{\cal Y}) =
-\frac{\partial Z^{(R_1,R_2)}}{\partial x}~~.
\end{equation}
For every orbit we can define the constant energy 
\begin{equation}\label{eq:Zhamilt}
Z^{(R_1,R_2)}(x,\tau;{\cal Y},Y_p)-Y_p \equiv 
\zeta^{(R_1,R_2)}(x,\tau;{\cal Y})= {\cal E}~~.
\end{equation}
Note that since $Y_p$ appears only as an additive constant in 
$Z^{(R_1,R_2)}$, the function $\zeta^{(R_1,R_2)}$ does not depend 
on $Y_p$. Also, according to \eqref{eq:fake_prop_ecc} and 
\eqref{eq:delau-garf-coord}, we have ${\cal Y}=\frac{e_{p,0}^2}{2}$. 
Then, for fixed value of ${\cal E}$, we can express $\tau$ as an 
explicit function of $x$, 
\begin{equation}\label{eq:taufunct}
\zeta^{(R_1,R_2)}(x,\tau;{\cal Y}) = {\cal E}  \quad \Longrightarrow 
\quad \tau = \tau({\cal E},x;{\cal Y})~~. 
\end{equation}
Replacing~\eqref{eq:taufunct} in~\eqref{eq:xdot}, we get
\begin{equation}\label{eq:xdot2}
\frac{\textrm{d}x}{\textrm{d}t} = f (x,\tau({\cal E},x;{\cal Y})
;{\cal Y}) \quad \Longrightarrow \quad \text{d}t = 
\frac{\text{d}x}{f (x,\tau({\cal E},x;{\cal Y}) ;{\cal Y})}~~,
\end{equation}
whereby we can derive an expression for the synodic period $T_{syn}$
\begin{equation}\label{eq:synper_int}
T_{syn} = \oint \frac{\text{d}x}
{f (x,\tau({\cal E},x;{\cal Y}) ;{\cal Y})}~~,
\end{equation}
and thus the synodic frequency
\begin{equation}\label{eq:synfreq}
\omega_{s} = \frac{2\pi}{T_{syn}}~~.
\end{equation}
In practice, \eqref{eq:taufunct} is hard to invert analytically, 
and hence, the integral (\ref{eq:synper_int}) cannot be  explicitly 
computed. We thus compute both expressions numerically on grids of 
points of the associated invariant curves on the plane $(\tau,x)$, 
or by integrating numerically \eqref{eq:xdot2} as a first order 
differential equation.

2) We now compute the fast and secular frequencies $\omega_f$, 
$g$. To compute $\omega_f$, we use the 
equation
\begin{equation}\label{eq:fastfreq1}
\omega_{f} = \, \frac{1}{T_{syn}} \int_{0}^{T_{syn}} 
\frac{\textrm{d}\phi_f}{\textrm{d}t} \, \textrm{d}t = 
\, \frac{1}{T_{syn}} \int_{0}^{T_{syn}} 
\frac{\partial Z^{(R_1,R_2)} (x,\tau;{\cal Y})}{ \partial {\cal Y}} \, 
\textrm{d}t~~.
\end{equation}
Replacing~\eqref{eq:xdot2} in~\eqref{eq:fastfreq1}, we generate an
explicit formula for the fast frequency
\begin{equation}\label{eq:fastfreq3}
\omega_{f} = \frac{1}{T_{syn}} \oint
\frac{1}{f (x,\tau({\cal E},x;{\cal Y}) ;{\cal Y})} \,
\frac{\partial Z_{R_1,R_2} (x,\tau({\cal E},x;{\cal Y});{\cal Y})}
{ \partial {\cal Y}}\, \textrm{d}x~~.
\end{equation}
Since $Z^{(R_1,R_2)}(x,\tau;{\cal Y},Y_p) = Y_p +
\zeta^{(R_1,R_2)}(x,\tau;{\cal Y})$ we find $\dot{\theta}=1$ implying
$g=1-\omega_f$.

All the frequencies are thus functions of the labels 
${\cal E}$ and ${\cal Y}$, which, in the integrable normal form 
approximation, label the proper libration and the proper eccentricity 
of the orbits. In the normal form approach one has $e_{p,0}=e_p=$~const, 
implying ${\cal Y}=e_p^2/2$. If, as in~\cite{PaezEfthy2015}, we fix a 
scanning line of initial conditions $x_{in} = B u_{in}=\tau_{in}-\tau_0$, 
with $B$ a constant, the energy ${\cal E}$, for fixed $e_p$, becomes a 
function of the initial condition $u_{in}$ only. Thus, $u_{in}$ 
represents an alternative label of the proper libration (see Section 3 
of~\cite{PaezEfthy2015} for a detailed discussion of this point). 
With these conventions, all three frequencies become functions of 
the labels $(u_{in},e_p)$. A generic resonance condition then reads
\begin{equation}\label{eq:rescond_functPhi}
\Phi_{m_f,m_s,m}(u) = m_f \omega_f(e_p,u_{in}) 
+ m_s \omega_s(e_p,u_{in})+ m g(e_p,u_{in})=0~~.
\end{equation}
For fixed resonance vector $(m_f,m_s,m)$, Eq.~\eqref{eq:rescond_functPhi} 
can be solved by root-finding, thus specifying the position of the 
resonance on the plane of the proper elements $(u_{in},e_p)$. 

\begin{figure}[t]
  \centering
  \includegraphics[width=.75\textwidth]{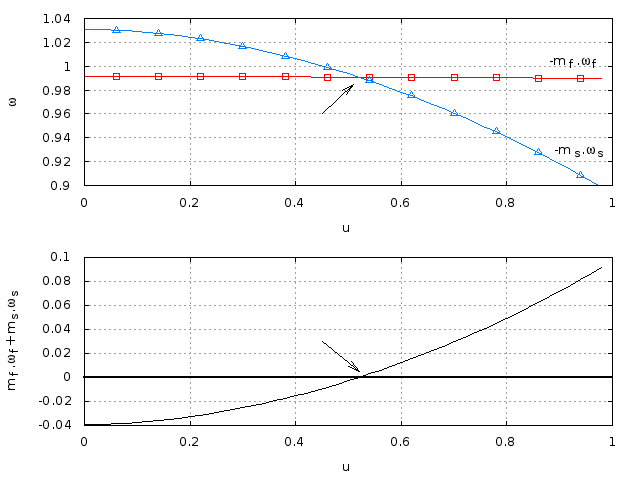}
  \caption{Representation of the evolution of the frequencies as
    function of $u$.  In the upper panel, $m_f \omega_f$ (red square
    points) and $-m_s\omega_s$ (blue triangle points).  In the
    lower panel, the evolution of the function
    $m_f\omega_f+m_s\omega_s$ (black curve). The arrows denote the
    point where the frequencies accomplish the resonant condition
    $m_f\omega_f+m_s\omega_s=0$, giving the position of the resonance
    in terms of $u$. For this example, we choose the resonance $1$:$8$,
    corresponding to $m_f=1$, $m_s=8$, $\mu=0.0024$, $e'=0.04$ and a
    representative value for $e_{p,0}=0.05$.}
  \label{fig:freq_fig}
\end{figure}

As an example, Fig.~\ref{fig:freq_fig}, shows $\omega_f$ and $\omega_s$,
as well as the function $\Phi_{1,8,0}(e_p,u_{in})$, as a function of 
$u_{in}$ for the parameters $\mu=0.0024$, $e'=0.04$ and a fixed 
value of $e_p=0.05$. The arrow in the lower panel marks the position 
of the resonance. Changing the value of $e_p$ in the same range as the 
one considered in our numerical FLI stability maps ($0<e_{p,0}<0.1$), 
we specify $u_{in}$ all along the locus of the resonance projected 
in the stability map. Repeating this computation for several transverse
resonances $(m_f,m_s,m)$, we are able to trace the location of each
of them.

\begin{figure}[h]
  \centering \includegraphics[width=.95\textwidth]{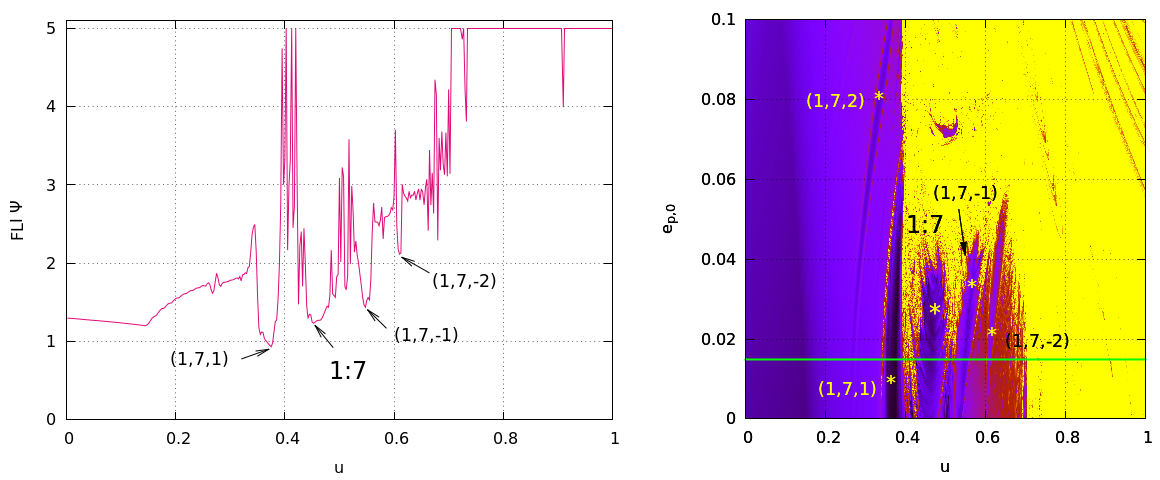} 
  \caption[Local minima of the FLI as tracers of the resonance center]{FLI
  $\Psi$ as function of $u$, for fixed
  parameters $\mu~=~0.0031$, $e'~=~0.04$ and $e_{p,0}~=~0.015$ (right
  panel). The local minima give a good approximation of the position
  of the centers of each resonance. The orbits whose corresponding FLI
  values are plotted in the left panel lie on the green line on top of
  the FLI map (right panel). The confirmation of each resonance is 
  done by frequency analysis.}  \label{fig:linesFLI.png}
\end{figure}

In order to test the accuracy of the above method, we compare the
results of the semi-analytical estimation with the position of the
resonances extracted from the FLI maps computed 
in~\cite{PaezEfthy2015}. Under the assumption that the local
minimum of the FLI in the vicinity of a resonance gives a good
approximation of the resonance center, we study the curves of the FLI
$\Psi$ as a function of $u$, for a
fixed value of $e_{p,0}$.  Figure~\ref{fig:linesFLI.png} gives an example
for $\mu=0.0031$, $e'=0.04$, $e_{p,0} = 0.015$, where we
choose four candidates as centers of the resonances $(1,7,1)$,
$1$:$7$, $(1,7,-1)$ and $(1,7,-2)$. We confirm the resonant character
 of these orbits also by performing a numerical Frequency
Analysis~\cite{Laskar-04}. By changing the value of $e_{p,0}$ along
the interval $[0,0.1]$, we can depict the centers of the resonances on
top of the FLI maps.

\begin{SCfigure}
  \centering \includegraphics[width=.60\textwidth]{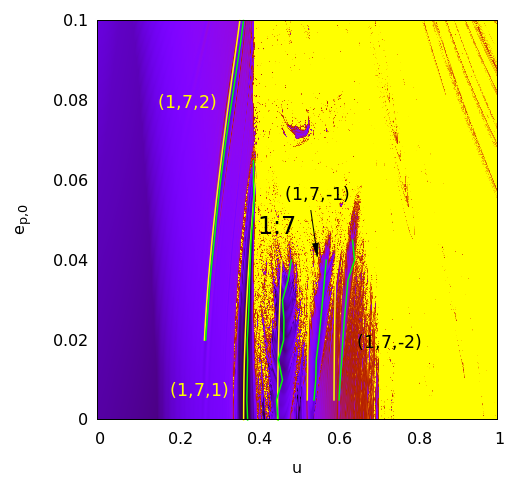} 
  \caption[Main
  and transverse secondary resonances located by $Z^{(R_1,R_2)}$
  and by FLI $\Psi$ minima - 1]{Main
  and transverse secondary resonances located by $Z^{(R_1,R_2)}$
  (yellow) and the estimation of FLI $\Psi$ minima (green).  In this
  example, $\mu=0.0031$, $e'=0.04$, $m_f=1$, $m_s=7$, $m=0,\pm 1,\pm
  2$.  Labels indicate the corresponding resonance in each
  case. \vspace{0.8cm}} 
  \label{fig:transverses1.png}
\end{SCfigure}

\begin{SCfigure}
  \centering 
  \includegraphics[width=.60\textwidth]{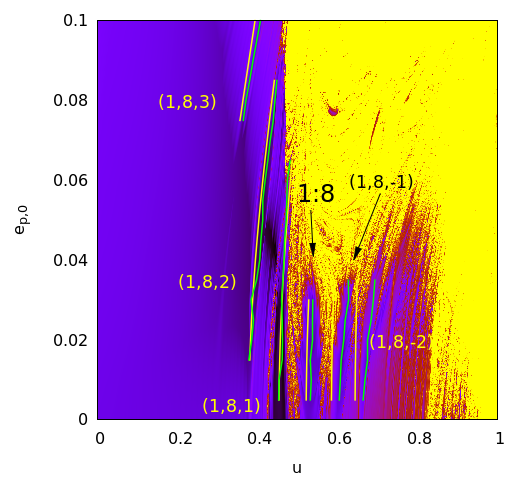} 
  \caption[Main
  and transverse secondary resonances located by $Z^{(R_1,R_2)}$
  and by FLI $\Psi$ minima - 2]{Same
  as Fig.~\ref{fig:transverses1.png}, for $\mu=0.0024$, $e'=0.06$, and
  $m_f=1$, $m_s=8$, $m=0,\pm 1,\pm 2,3$.  
  \vspace{1.2cm}} 
  \label{fig:transverses2.png}
\end{SCfigure}

\begin{SCfigure}
  \centering
  \includegraphics[width=.60\textwidth]{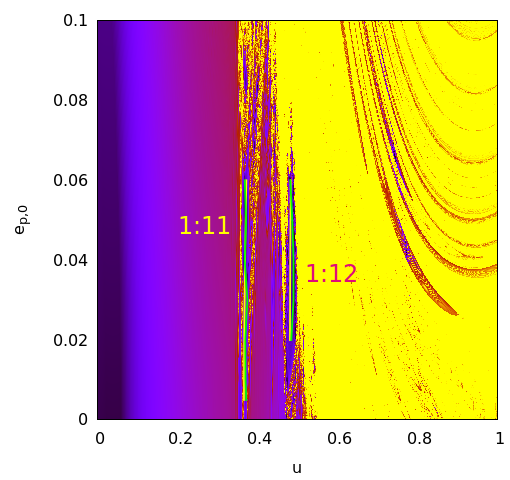}
  \caption[Main
  and transverse secondary resonances located by $Z^{(R_1,R_2)}$
  and by FLI $\Psi$ minima - 3]{Same
  as Fig.~\ref{fig:transverses1.png}, for $\mu=0.0014$, $e'=0.02$,
  and  $m_f=1$, $m_s=11,12$, $m=0$.
  \vspace{1.2cm}} 
  \label{fig:transverses3.png}
\end{SCfigure}

\begin{table}[h]
\begin{center}
\begin{tabular}{|c c c c c|}
\hline
Resonance & $\mu$, $e'$ & $\overline{u_{{\cal Z}}}$ & $\overline{u_{\Psi}}$
& $\overline{\delta u_{in}}$ \\
\hline
$1$:$7$   & $0.0031$, $0.04$ & 0.453908 & 0.463308  &  2.129422$\snot[-2]$ \\
$(1,7,1)$ &      $''$       & 0.377456 & 0.380947  &  1.417910$\snot[-2]$ \\
$(1,7,2)$ &      $''$       & 0.306036 & 0.312011  &  1.880279$\snot[-2]$ \\
$(1,7,-1)$&      $''$       & 0.527218 & 0.554430  &  4.885329$\snot[-2]$ \\
$(1,7,-2)$&      $''$       & 0.593373 & 0.618057  &  3.964370$\snot[-2]$ \\
$1$:$8$   & $0.0024$, $0.06$ & 0.524485 & 0.535153  &  1.993063$\snot[-2]$ \\
$(1,8,1)$ &      $''$       & 70.465475 & 0.464924  &  6.377401$\snot[-3]$ \\
$(1,8,2)$ &      $''$       & 0.406439 & 0.412246  &  1.605145$\snot[-2]$ \\
$(1,8,3)$ &      $''$       & 0.374879 & 0.385020  &  2.617987$\snot[-2]$ \\
$(1,8,-1)$&      $''$       & 0.587834 & 0.616093  &  4.572688$\snot[-2]$ \\
$(1,8,-2)$&      $''$       & 0.646464 & 0.679154  &  4.796435$\snot[-2]$ \\
$1$:$11$  & $0.0014$, $0.02$ & 0.367663 & 0.370842  &  9.264243$\snot[-3]$ \\
$1$:$12$  &      $''$       & 0.482117 & 0.486631  &  1.021940$\snot[-2]$ \\
\hline
\end{tabular}
\end{center}
\caption{Averaged values of $u_{{\cal Z}}$, $u_{{\Psi}}$ and $\delta u_{in}$ for the resonances in Figures~\ref{fig:transverses1.png}, 
\ref{fig:transverses2.png} and~\ref{fig:transverses3.png}}
\label{tab:errors}
\end{table}

Figures~\ref{fig:transverses1.png},~\ref{fig:transverses2.png}
and~\ref{fig:transverses3.png} show examples of these computations,
for the parameters $\mu=0.0031$ and $e'=0.04$, $\mu=0.0024$ and $e'=0.06$, 
$\mu=0.0014$ and $e'=0.02$, respectively. The normal form predictions are
superposed as yellow lines upon the underlying FLI stability maps 
while the centers of each resonance, as
extracted from the FLI maps, are denoted by the green
curves. Due to the numerical noise in the FLI curves, it is not possible to
clearly extract the position of the resonance centers for all values
of $e_{p,0}$, while a semi-analytic estimation (with varying levels of
accuracy) is always possible.  At any rate, in 
Figs.~\ref{fig:transverses1.png}-\ref{fig:transverses3.png},
we compare the position of the resonaces only
in these cases when both methods provide clear
results. Table~\ref{tab:errors} summarizes the results for the
location of the centers ($u_{{\cal Z}}$, $u_{\Psi}$) and the relative
errors ($\delta u_{in} = \frac{|u_{{\cal Z}} - u_{\Psi}|}{u_{\Psi}}$),
on average, for the resonances shown in the corresponding figures.

Regarding the overall performance of the estimation, we can
note that the level of approximation is very good for relatively low
values of $\mu$, $e_p$ and $u_{in}$, while the error in the predicted
position of the resonance increases to a few percent for greater
values of those parameters, with an upper (worst) value $6\%$ (see
Table~\ref{tab:errors}). This is the expected behavior for a normal
form method, whose approximation becomes worse with higher values of
the method's small parameter(s). Independently of this fact, the
normal form approach is based on the use of the basic model $H_b$ as a
starting Hamiltonian. This confirms that the basic Hamiltonian is able
to well approximate the fast and synodic dynamics of the
ERTBP. Additionally, the fact that we do not consider expansions in
terms of $\tau$ allows to retain accurate information about higher
order harmonics.  Finally, by using the relation between the fast
action $Y_f$ and the secular action $Y_p$, it is possible to estimate,
via $H_b$, the value of the secular frequency $g$, and, hence, to
determine also the position of transverse resonances in the plane of
proper element, even though these resonances have no 'width' in the
dynamics under the $H_b$.

\section{Conclusions}
Our main results in this work can be summarized as follows:

1) We have demonstrated the efficiency of the normal form approach
introduced in~\cite{PaezLocat2015} in order to determine
the position of resonances in the space of proper elements in the
tadpole domain of Trojan motions. As discussed in Section 1, the main
advantage of the new approach is based on avoiding to perform series
expansions with respect to the synodic co-ordinates around the
Lagrangian equilibrium points $L_4$ and $L_5$. The latter expansions
are subject to a poor convergence.  On the contrary, the method
proposed here circumvents the issue of this poor convergence, and even
relatively low order expansions can give results accurate down to an
error of a few percent only.

2) We have applied the above normal formal approach in a Hamiltonian
model called `the basic model' in~\cite{PaezEfthy2015}. This is a
model allowing to efficiently separate the secular part of the
Hamiltonian from the part representing the dynamics in the fast and
synodic degrees of freedom.  We should emphasize here that in the case
of the 1:1 Mean Motion resonance this separation is non-trivial and
proceeds along different lines than in the case of other mean motion
resonances. This is due to the non-trivial nature of the forced
equilibrium at the 1:1 MMR. Yet, as detailed in Section 2 above, the
`basic model' allows to study the dynamics in the fast (${\cal O}(1)$)
and intermediate (${\cal O}(\sqrt{\mu})$ frequency scales in a unified
way independently of the number of the primary disturbing bodies in
the system. As shown in Section 3, normalizing the basic model turns
to be sufficient for most analytical predictions regarding the
dynamics in these timescales.

The present methods can be easily adapted in two cases: i) considering 
Trojan motions off the plane (spatial ERTBP or RMPP), and ii) considering 
a time-varying configuration of the $S$ primaries, beyond the 
quasi-periodic secular variations of Eq.~\eqref{eq:h_rmpp}. For the long 
term stability, as well as the possibility of captures or escapes of 
small Trojan bodies (asteroids and/or hypothetical exo-planets), 
in~\cite{RobBod-09} the authors demonstrated that a crucial role is 
played by resonances crossing the Trojan domain during the phase of 
planetary migration. In this case, it would be desirable to be able 
to specify the time-varying locus of the secondary resonances via 
analytical techniques. Let us note here that the depletion rate of 
a Trojan swarm along secondary resonances is, in principle, related 
to the size of the remainder function of the normal form proposed in 
Section 3. In simple Hamiltonian models, it has been found that 
the diffusion rate goes as a power-law of the size of the remainder 
function (see~\cite{Efthy-08},~\cite{EftHar-13}). The degree up to which 
such laws are applicable in a physical context like the co-orbital 
resonance is unknown, and this question poses a possible extension 
of the present work. 

\vspace{0.5cm}
\noindent
{\bf Acknowledgements:} During this work, R.I.P. was supported by the
Astronet-II Marie Curie Training Network (PITN-GA-2011-289240) and by
the project ``Dynamics of the celestial bodies in the neighborhood of
the Lagrangian points'' of the University of Rome ``Tor Vergata''.

\section*{Appendix}
Variables corresponding to the three degrees of freedom appearing in the
expression of the Basic Hamiltonian $H_b$ in Eq.\eqref{eq:hbasic}, 
$(u,v)$, $(Y_f,\phi_f)$ and $(Y_p,\phi_p)$, in terms of the orbital elements:

\begin{equation}
u = \lambda - \lambda' - \frac{\pi}{3}~~,
\end{equation}

\begin{equation}
v = \sqrt{a} - 1~~,
\end{equation}

\begin{displaymath}
\beta = \omega - \phi'~~, 
\end{displaymath}

\begin{displaymath}
y = \sqrt{a} \left( \sqrt{1-e^2} -1 \right)~~,
\end{displaymath}

\begin{displaymath}
V = \sqrt{-2y} \sin \beta - \sqrt{-2y_0} \sin \beta_0~~,
\end{displaymath}

\begin{displaymath}
W = \sqrt{-2y} \cos \beta - \sqrt{-2y_0} \cos \beta_0~~,
\end{displaymath}

\begin{displaymath}
Y = - \left( \frac{W^2 + V^2}{2} \right)
\end{displaymath}

\begin{equation}
\phi = \arctan \left( \frac{V}{W} \right) 
\end{equation}

\begin{equation}
\phi_f = \lambda' - \phi~~,
\end{equation}

\begin{equation}
Y_f = \int \frac{\partial E}{\partial \lambda'} \mathrm{d} t + v~~,
\end{equation}

\begin{equation}
Y_p = Y - Y_f~~,
\end{equation}
where $\lambda$, $\omega$, $a$ and $e$
are the mean longitude, the longitude of the perihelion,
the major semiaxis and eccentricity of the Trojan body, 
$\lambda'$ and $\phi' = \omega'$ are the
mean longitude and longitude of the perihelion of the
perturber, $\beta_0 = \pi/3$, $y_0 = \sqrt{1-e'^2} -1$,
and $E$ represents the total energy of the Trojan as 
computed from Eq.~\eqref{eq:h_rmpp} (see \cite{PaezEfthy2015} for
further details in the construction).


\end{document}